\documentclass[reprint,aps,apl,superscriptaddress,floatfix]{revtex4-2}

\usepackage{amsmath,amssymb,bm}
\usepackage{graphicx}
\usepackage{hyperref}
\usepackage{booktabs}

\DeclareMathOperator*{\RMS}{RMS \,}
\DeclareMathOperator*{\med}{median \,}

\begin{document}

\title{Low-rank modal endpoints from beat-resolved retinal arterial Doppler holography velocity waveforms characterize the response to flicker provocation}

\author{Sienna O'Shea}
\affiliation{Institut Langevin, CNRS, ESPCI Paris, PSL University, Paris, France}

\author{Maxime Boy Arnould}
\affiliation{Institut Langevin, CNRS, ESPCI Paris, PSL University, Paris, France}

\author{Adrien Gordon}
\affiliation{Institut Langevin, CNRS, ESPCI Paris, PSL University, Paris, France}

\author{Yann Fischer}
\affiliation{Institut Langevin, CNRS, ESPCI Paris, PSL University, Paris, France}

\author{Zacharie Auray}
\affiliation{Institut Langevin, CNRS, ESPCI Paris, PSL University, Paris, France}

\author{Michael Atlan}
\affiliation{Institut Langevin, CNRS, ESPCI Paris, PSL University, Paris, France}

\date{\today}

\begin{abstract}
Conventional retinal flicker endpoints quantify changes in diameter, mean velocity, or flow but do not capture the modal concentration of the cardiac velocity waveform during neurovascular stimulation. Here, beat-resolved retinal Doppler holography and low-rank modal decomposition are combined to derive compact arterial endpoints from unfiltered arterial segment-velocity waveforms sampled across beats and vessel locations. For each acquisition, locally centered waveforms are assembled into a common matrix and decomposed by singular-value decomposition (SVD). Robust summaries quantify total pulsatile scale, modal amplitude, residual amplitude, mean-to-pulsatile balance, and singular-spectrum dimensionality; the same construction can also be applied descriptively to individual beats. We demonstrate the framework using 33 temporally ordered Baseline~1/Flicker/Baseline~2 acquisitions from one eye. Relative to the pooled baselines, 13-Hz flicker was associated with lower centered-waveform RMS scale $R_0$ and leading-mode amplitude $A_1$, and with higher robust residual-amplitude ratios $\rho_1$ and $\rho_2$, mean-to-pulsatile ratio MPR, effective rank, and participation ratio; these contrasts showed exploratory within-session separation after Holm adjustment. No evidence of differences in the absolute residual amplitudes $R_1$ and $R_2$ was detected; together, these observations are consistent with reduced concentration in the leading acquisition-specific arterial pulse mode rather than an increase in absolute residual pulsatility.
\end{abstract}

\maketitle

\section{Introduction}
\label{sec:introduction}

Neurovascular coupling matches local blood supply to neuronal activity. The retina is an optically accessible part of the central nervous system and therefore provides a direct, non-invasive setting in which to study this process. Flickering light increases retinal neural activity and evokes functional hyperemia through coordinated responses of the arteriolar--capillary--venular network.\cite{Newman2013} In humans, flicker-induced increases in retinal arterial velocity and flow have been demonstrated with laser Doppler methods and Doppler optical coherence tomography (OCT).\cite{Garhofer2004FlickerFlow,Wang2011DopplerOCT,Aschinger2017TotalFlow}

Flicker-evoked retinal vascular responses are impaired or otherwise dysregulated in several major microvascular and neurodegenerative disorders. In diabetes, attenuated vasodilation has been detected before clinically visible retinopathy, becomes more pronounced with disease severity, is already measurable in prediabetes, and has been associated prospectively with retinopathy progression.\cite{Garhofer2004Diabetes,Mandecka2007Diabetes,Lott2013Prediabetes,Lim2017DRProgression} Blunted retinal responses have also been associated with systemic endothelial dysfunction and reported in glaucoma and chronic ischemic white-matter disease; in the latter, retinal reactivity was associated with middle-cerebral-artery vasoreactivity.\cite{Pemp2009Endothelial,Gugleta2013Glaucoma,Bettermann2012WhiteMatter} Complementary CO$_2$-provocation measurements in cerebral small-vessel disease have further linked retinal vascular reactivity to white-matter-hyperintensity burden and cerebral BOLD reactivity.\cite{Blair2025cSVD} Findings across the Alzheimer spectrum are more heterogeneous, ranging from reduced response magnitude to increased but delayed and abnormally prolonged dilation.\cite{Querques2019Alzheimer,Kotliar2017Alzheimer} This heterogeneity suggests that peak dilation alone may be insufficient: response timing, recovery, full-waveform dynamics, and modal pulsatile concentration may provide complementary disease-sensitive information beyond conventional peak or mean-flow endpoints.

Retinal flicker provocation has consequently become a practical assay of vascular reactivity. Dynamic vessel analysis measures diameter changes in selected retinal vessels; Doppler OCT estimates velocity and total flow; laser speckle flowgraphy follows rapid flow-index changes at the optic nerve head and peripapillary vessels; and OCT angiography extends the assessment to depth-resolved and capillary-scale networks.\cite{Aung2024,Kallab2021PlexusSpecific,Huang2025CircularOCTA} Full-field and adaptive-optics measurements further resolve parafoveal flow and vessel-wall dynamics.\cite{Warner2020ParafovealFlow,Senee2025SciAdv} This methodological diversity is clinically relevant because altered retinal neurovascular coupling has been reported in diabetes and glaucoma and is being investigated as a retinal marker of cerebral and systemic microvascular dysfunction.\cite{Kwan2020,Liu2023fOCTA,Gugleta2013Glaucoma,Peterfi2024}

The measurement remains challenging. Flicker-response magnitude and kinetics depend on stimulus type, duration, spatial extent, adaptation state, and analysis window.\cite{Aung2024,Senee2025SciAdv} Moreover, the small evoked vascular changes are superimposed on heartbeat-driven pulsatility and slower spontaneous vasomotion. Continuous high-speed measurements can distinguish these components more reliably than sparse pre--post sampling, while recent wall-resolved imaging shows that response amplitude, timing, and baseline variability provide complementary information.\cite{Senee2025SciAdv} Test--retest studies likewise show that standardized acquisition and stable quantitative endpoints are prerequisites for comparisons across sessions and cohorts.\cite{Kalitzeos2026Repeatability}

Most established endpoints reduce the flicker response to a change in vessel diameter, mean velocity, total flow, or a flow surrogate, together with its peak or time course. These quantities remain physiologically important, but they do not describe the modal concentration of the cardiac waveform itself. The pulse waveform is commonly treated as background variability to be averaged out, although it encodes the interaction between cardiac forcing, vascular tone, compliance, resistance, and downstream runoff. A flicker response may therefore change pulsatile scale, alter the expression of a shared pulse shape, or reduce concentration in its leading mode even when mean-flow endpoints alone provide an incomplete description. The arterial waveform provides a direct view of this inflow response.

Laser Doppler holography provides non-invasive, full-field retinal blood-flow measurements with sufficient temporal resolution to resolve cardiac waveforms simultaneously at multiple arterial locations.\cite{Puyo2018InVivoLDH,Puyo2019Waveform} The resulting data span cardiac phase, beats, and vessel locations and are naturally suited to low-rank analysis. Temporal singular-value decomposition (SVD) modes represent waveform patterns shared across an acquisition, their scores quantify the expression of those patterns at individual beat--location samples, and residuals measure structure not explained by the leading modes.

A complementary shape-first library has defined beat-wise LDH descriptors that are invariant or robust to a positive multiplicative velocity gain, with the aim of transportability across sessions, instruments, and sites when absolute calibration is uncertain.\cite{Buisson2026Transportable} The present low-rank endpoints complete that framework at the acquisition and beat levels by adding measures of pulsatile scale, shared-mode expression, residual structure, and spectral dimensionality. Their normalized members retain the same transportability objective, whereas the absolute-amplitude members preserve calibration-sensitive physiological scale.

A preliminary analysis of a separate dataset from one healthy volunteer showed a coupled arterial response to 13-Hz flicker: leading-mode pulsatile amplitude decreased while the rank-1 residual-amplitude ratio increased relative to the flanking baseline epochs.\cite{Doucet2026LowRank} This suggested that flicker changes both waveform scale and the relative residual amplitude not captured by the leading acquisition-specific arterial pulse mode.

Here we convert these high-dimensional waveform fields into compact endpoints for total centered-waveform amplitude, modal amplitude, residual amplitude, robust residual-amplitude ratio, mean-to-pulsatile balance, and spectral dimensionality. The construction uses all valid beat--location waveforms and robust median summaries, reducing dependence on a single vessel, location, beat, or manually selected waveform feature. Each endpoint is directly calculable at the acquisition or beat level and is therefore compatible with automated quality control, longitudinal studies, and patient-cohort analysis. The present results are restricted to repeated Baseline~1/Flicker/Baseline~2 measurements from one eye. They demonstrate the arterial endpoint framework and its within-session separation, but do not establish reproducibility across eyes or individuals.

\section{Experimental methods}
\label{sec:experimental_methods}

\subsection{Optical system}

\begin{figure}[htbp]
\centering
\includegraphics[width=\linewidth]{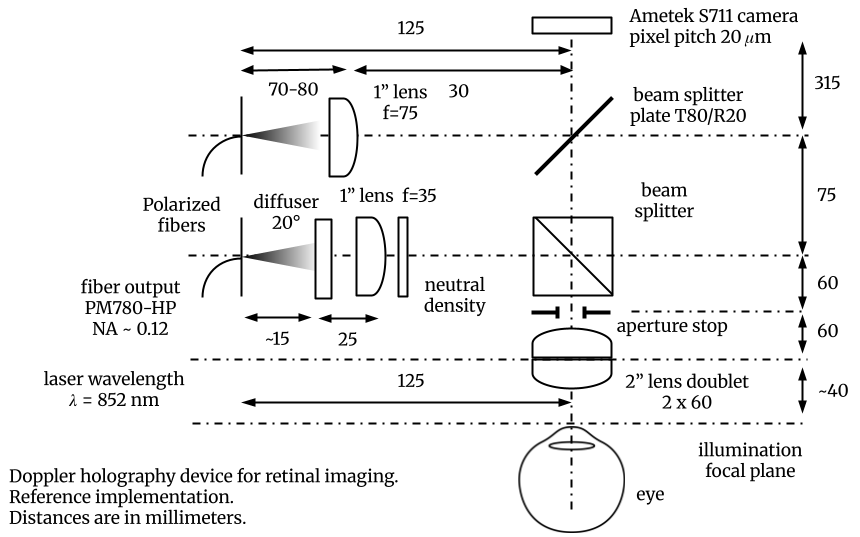}
\caption{Clinical laser Doppler holography prototype and optical layout. Near-infrared light is split into illumination and reference arms of an inline Mach--Zehnder interferometer. The diffused illumination beam provides safe wide-field retinal illumination, and the cross-polarized backscattered field is recombined with the reference beam on a high-speed camera. Numerical processing reconstructs the retinal Doppler images and estimates the ocular wavefront.}
\label{fig:setup}
\end{figure}

Retinal interferograms were acquired with an on-axis digital holography ophthalmoscope based on an inline Mach--Zehnder interferometer operating at $\lambda=852$~nm. Light from a diode laser (Thorlabs FPV852P) was divided between the illumination and reference arms with a $90/10$ power ratio using polarization-maintaining fibers. In the illumination arm, an engineered diffuser (Thorlabs ED1-C20-MD, $20^\circ$ circular top-hat) and two 60-mm biconvex lenses provided diffuse Maxwellian illumination.\cite{Bratasz2022Diffuse} Cross-polarized light backscattered by the fundus was recombined with the reference beam and recorded with a Phantom S711 camera over $512\times320$ pixels with a $20~\mu\mathrm{m}$ pixel pitch.\cite{Auray2025UltrahighSpeed}

\subsection{Recording conditions}

Retinal holograms were acquired without pharmacological pupil dilation in one eye of a healthy 25-year-old man after written informed consent. During acquisition, the participant viewed a static fixation target with the contralateral eye. The procedures adhered to the Declaration of Helsinki and were authorized by the local ethics boards (CPP Sud-Est~III: 2019-021B; ANSM/IDRCB: 2019-A00942-55; ClinicalTrials.gov: NCT04129021). The optical power at the diffused focal spot entering the eye was 5.0~mW. In the reported assessment of this 852-nm diffuse Maxwellian configuration, the most conservative design recommendation was an operating power not exceeding approximately 6~mW, set by the ISO 15004-2:2024 iris estimate; the ANSI Z80.36-2021 retinal design estimate was approximately 11~mW, and the anterior-segment condition was not limiting.\cite{Bratasz2022Diffuse}

\subsection{Velocity assessment}

Each acquisition contained 163,840 interferograms recorded over approximately 4.4~s at approximately 37~kHz.\cite{Auray2025UltrahighSpeed} The complex optical field was reconstructed by Fresnel propagation over 480~mm. Doppler processing used non-overlapping 256-frame temporal Fourier windows and removal of dominant SVD components associated with clutter and bulk motion, yielding approximately 145 velocity maps per second. Following the published velocity-assessment procedure,\cite{Fischer2024RetinalFlow} the normalized spectral second moment over $f_1=6$~kHz to $f_2=18.5$~kHz was calculated for each vessel location as
\begin{equation}
M_2
=
\frac{\displaystyle\int_{f_1}^{f_2} f^2 S(f)\,\mathrm{d}f}
{\displaystyle\int_{f_1}^{f_2} S(f)\,\mathrm{d}f},
\label{eq:normalized_second_moment}
\end{equation}
where $S(f)$ is the local Doppler power spectrum; $M_{2,\mathrm{bkg}}$ was calculated analogously from the local tissue-background spectrum. The signed differential RMS spectral broadening was defined as
\begin{equation}
\Delta f
=
\operatorname{sgn}\!\left(M_2-M_{2,\mathrm{bkg}}\right)
\sqrt{\left|M_2-M_{2,\mathrm{bkg}}\right|},
\label{eq:doppler_frequency}
\end{equation}
and the LDH velocity calibration used here was
\begin{equation}
v
=
\frac{\lambda\Delta f}{\mathrm{NA}}.
\label{eq:doppler_velocity}
\end{equation}
The resulting arterial velocity waveforms were organized as $v(t,b,k,r)$, where $t$ is within-beat time, $b$ indexes beats, and $(k,r)$ identifies the vessel location. Here, $\lambda=852$~nm and the numerical aperture of the eye is $\mathrm{NA}=0.124$. Velocity was expressed in $\mathrm{mm\,s^{-1}}$.

\subsection{Data and experimental epochs}
\label{sec:dataset}

\begin{figure}[t]
\centering
\includegraphics[width=0.48\linewidth]{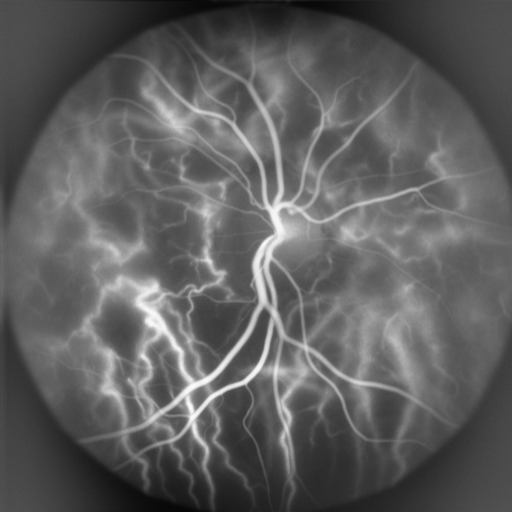}
\includegraphics[width=0.48\linewidth]{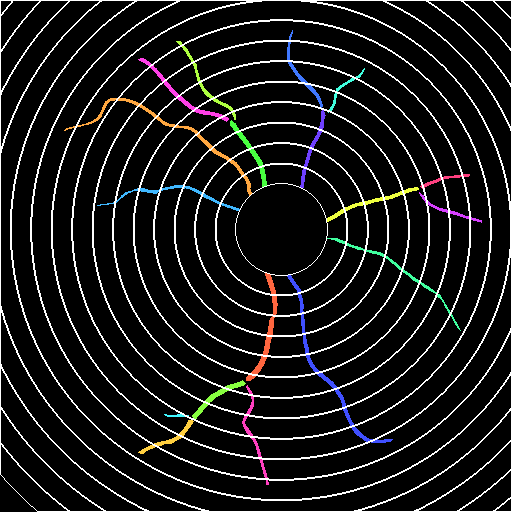}
\includegraphics[width=\linewidth]{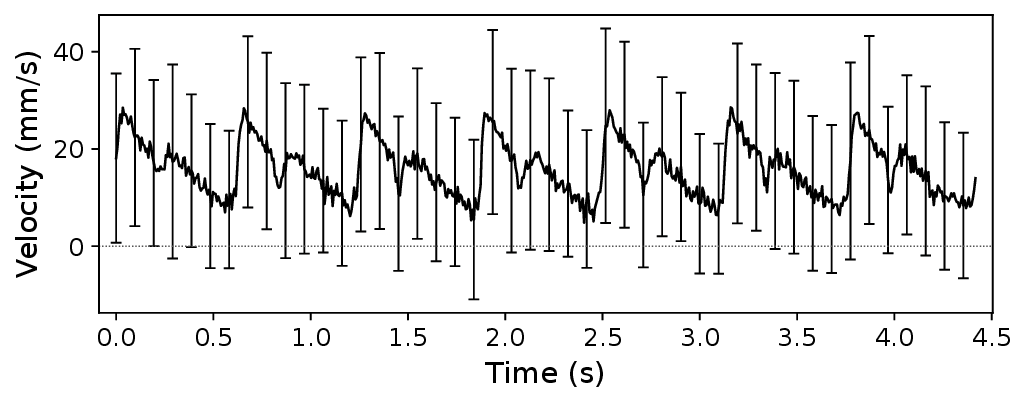}
\caption{Representative arterial data from one Baseline~2 acquisition of the analyzed eye.
Top left: frequency map. Top right: automatically identified arterial branches and optic-disc-centered candidate annuli used to define the branch--radius sampling locations.
Bottom: full-acquisition arterial segment velocity versus wall-clock time (s).
The black curve is the spatial mean over segment locations $(k,r)$;
whiskers show $\pm$ one sample spatial standard deviation, drawn every $0.1\,\mathrm{s}$.}
\label{fig:frequency_velocity_overview}
\end{figure}

Results are presented for one eye recorded with a Baseline~1/Flicker/Baseline~2 protocol. The dataset contains 33 acquisitions: nine during Baseline~1, fourteen during 13-Hz flicker, and ten during Baseline~2.

The flicker stimulus was delivered by a white LED as a 13-Hz square wave with a 50\% duty cycle. Its optical power was approximately $0.2$~mW at the corneal plane, measured within the $\varnothing 9.5$-mm circular aperture of a Thorlabs PM16-120 power sensor. The adaptation period, defined as the delay between flicker onset and the start of an acquisition, was $\gtrsim 5$~s. Flicker remained active throughout each 4.4-s flicker acquisition. It was switched off for most of the approximately 30-s inter-acquisition interval, during which the participant rested with both eyes closed, and was restarted at least 5~s before each subsequent flicker acquisition.

The principal retinal vessels used for sampling were first delineated from the averaged power-Doppler image using a U-Net-based binary segmentation; artery--vein classification then combined the power-Doppler map with cardiac-pulse-derived temporal cues, including pixelwise correlation to the arterial pulse and systolic--diastolic contrast images.\cite{Dubosc2026Segmentation} Downstream segmentation and analysis used these upstream masks and labels without an interactive segmentation-correction step; thus, a location's validity in the low-rank analysis did not encode a separate manual segmentation decision.

\subsection{Waveform validity and quality control}

For the low-rank analysis, we used the raw per-beat arterial segment-velocity waveforms, not the harmonic-band-limited waveform product. Systolic landmarks were detected from the average arterial velocity waveform over all segmented retinal arteries: after 15-Hz low-pass filtering, peaks of the temporal derivative above its 95th percentile and separated by at least 0.5~s were identified. Cardiac phase zero was the maximum detected systolic acceleration. A complete beat was an interval bounded by two consecutive validated systolic landmarks lying within the recorded signal. Leading and trailing partial cycles were therefore discarded, and an interval was retained only when its boundaries were strictly increasing and its beat period was finite and positive. Each retained beat was Fourier-resampled to $N_t=128$ within-beat samples.

The radial sampling geometry comprised 25 candidate annuli whose center positions translated with the acquisition-specific optic-disc center. These annuli defined the initial candidate geometry, whereas the upstream cross-section export retained only radial slots represented after vessel-intersection and signal-availability processing. Across acquisitions, approximately 15 arterial branches, 79 valid branch--radius locations, and seven complete beats were automatically identified. A branch--radius waveform in a retained beat was accepted when the upstream cross-section extraction produced finite velocity values at at least 95\% of its 128 phase samples. Locations below this threshold were omitted. For an accepted location, any remaining non-finite samples were replaced by that waveform's temporal mean before centering and hence became zero in the centered waveform. At least three valid waveform columns were required for an SVD. Dominant spatial SVD components were removed upstream to suppress clutter and bulk motion, but the low-rank implementation applied no additional automated residual-motion threshold; motion that remained finite was therefore not reclassified by the validity mask. No acquisition-level residual-motion exclusion is encoded in the 33-acquisition analysis reported here.

Figure~\ref{fig:frequency_velocity_overview} illustrates the arterial branch and radial sampling locations. In the representative processed acquisition, 16 of the 25 candidate radial slots were present in the upstream export. The acquisition contained seven complete beats, 15 branch slots, and 79 valid branch--radius locations per beat after waveform-validity filtering. Its joint acquisition matrix therefore contained $N_j=7\times79=553$ valid waveform columns, giving $q=\min(128,553)=128$. A representative single beat contained $N_{j,b}=79$ valid locations and consequently had $q_b=\min(128,79)=79$ singular values.

Beat period $T(b)$ is tracked as a potential physiological confound. The local temporal mean $\mu(b,k,r)$ is also retained because it reports the non-centered velocity level and may reveal acquisition drift or a sustained response not represented by the centered waveform. Each acquisition contributes one arterial endpoint vector. Beats and vessel locations contribute to these summaries but are not treated as independent observations in the acquisition-level analysis.

\section{Modal decomposition of beat-resolved waveforms}
\label{sec:model}

The derivation below is written for the arterial waveform field $v$.

\subsection{Waveform centering and matrix construction}

The local temporal mean is
\begin{equation}
\mu(b,k,r)
=
\left\langle v(t,b,k,r)\right\rangle_t,
\label{eq:mu_def}
\end{equation}
and the corresponding zero-mean waveform is
\begin{equation}
w(t,b,k,r)
=
v(t,b,k,r)-\mu(b,k,r).
\label{eq:w_def}
\end{equation}
Thus, $w$ retains the pulsatile waveform together with any zero-mean physiological or measurement variability, while $\mu$ is analyzed separately.

Let $j=1,\ldots,N_j$ index the valid beat--location combinations, with $j\leftrightarrow(b_j,k_j,r_j)$. At discrete within-beat times $t_n$, define
\begin{equation}
X_{nj}
\equiv
w_j(t_n)
\equiv
w(t_n,b_j,k_j,r_j).
\label{eq:X_stack}
\end{equation}
The waveforms are arranged columnwise as
\begin{equation}
\bm X
=
\begin{bmatrix}
w_1(t_1) & \cdots & w_{N_j}(t_1)\\
w_1(t_2) & \cdots & w_{N_j}(t_2)\\
\vdots & \ddots & \vdots\\
w_1(t_{N_t}) & \cdots & w_{N_j}(t_{N_t})
\end{bmatrix}
\in
\mathbb{R}^{N_t\times N_j}.
\label{eq:beat_location_matrix}
\end{equation}
Each column is one centered beat-resolved waveform, and each row contains samples at the same normalized cardiac phase across all valid beat--location combinations.

\subsection{SVD modes, scores, and residuals}
\label{sec:svd_construction}

The acquisition-level arterial SVD is
\begin{equation}
\bm X
=
\bm U\bm\Lambda\bm V^\top,
\label{eq:svd}
\end{equation}
where $\bm\Lambda$ contains the singular values
\begin{equation}
\lambda_1\geq\lambda_2\geq\cdots\geq\lambda_q\geq0,
\qquad
q=\min(N_t,N_j).
\label{eq:singular_value_order}
\end{equation}
The $m$th temporal mode and its score for waveform $j$ are
\begin{equation}
u_m(t_n)=U_{nm},
\qquad
a_m(j)=\lambda_m V_{jm}.
\label{eq:mode_and_score}
\end{equation}
For $M\leq q$, the centered waveform is represented as
\begin{equation}
w_j(t_n)
=
\sum_{m=1}^{M}a_m(j)u_m(t_n)
+r_M(t_n,j),
\label{eq:rank_M_waveform_model}
\end{equation}
where $r_M$ is the residual after removal of the first $M$ modes. After mapping $j$ back to $(b,k,r)$, the original waveform is
\begin{equation}
\begin{aligned}
v(t,b,k,r)
&=
\mu(b,k,r)\\
&\quad
+\sum_{m=1}^{M}a_m(b,k,r)u_m(t)\\
&\quad
+r_M(t,b,k,r).
\end{aligned}
\label{eq:waveform_model}
\end{equation}

The sign of each SVD mode is arbitrary. We orient each mode so that its median score is positive. Whenever $\med_j a_m(j)<0$, both $u_m$ and $a_m$ are multiplied by $-1$, leaving the reconstructed component unchanged.

Here, ``low-rank'' denotes analysis of the leading modes and their residuals, not a claim that the waveform matrix is accurately represented by rank one or two; the reported $\rho_1$, $R_{\mathrm{eff}}$, and PR values show that appreciable structure remains beyond the leading modes.

\begin{figure*}[t]
\centering
\includegraphics[width=\linewidth]{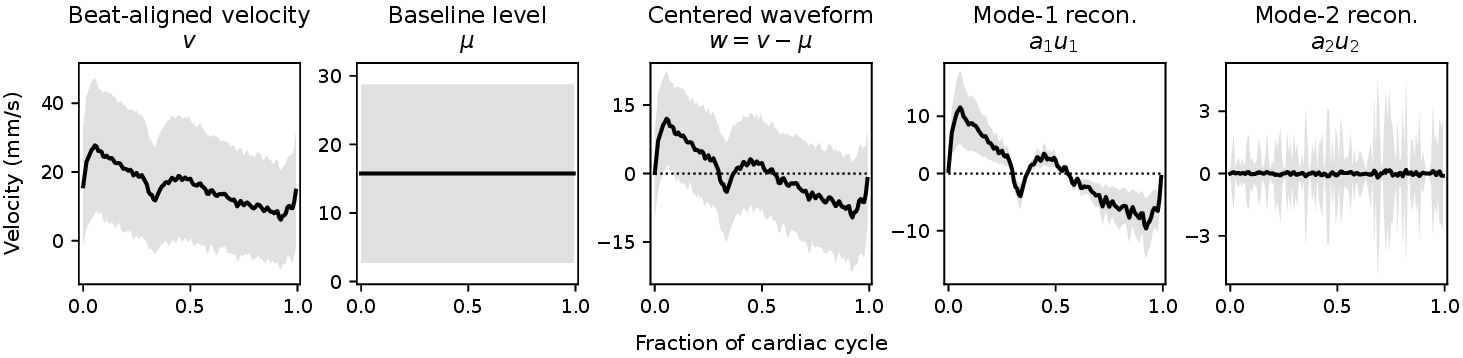}
\caption{Arterial waveform decompositions for a representative Baseline~2 acquisition from the analyzed eye (joint SVD).
Left to right: beat-aligned velocity $v$, temporal baseline $\mu(b,k,r)=\langle v\rangle_t$,
centered waveform $w=v-\mu$, and the joint mode-1 and mode-2 reconstructions $a_1u_1$ and $a_2u_2$.
The horizontal axis is cardiac phase as a fraction of the cycle:
each beat is resampled to $N_t=128$ samples with $\tau_n=n/N_t$ for $n=0,\ldots,N_t-1$, and $\tau=0$ is the maximum detected systolic acceleration.
In each panel the black curve is the mean over all valid beat--branch--radius columns $(b,k,r)$ at that phase,
and the gray band is mean $\pm$ one sample standard deviation over the same columns
(EyeFlow cross-column summary of the joint waveform field).
The $\mu$ panel is flat in phase because $\mu$ does not vary within a beat; its band reflects spread of $\mu$ across $(b,k,r)$ only.
Each panel has its own y-axis limits (mm/s); the $w$, $a_1u_1$, and $a_2u_2$ panels are centered on zero (dotted line).}
\label{fig:waveform_decomposition}
\end{figure*}

\section{Acquisition-level endpoints}
\label{sec:endpoints}

All endpoints in this section are calculated from the arterial decomposition. The joint acquisition-level SVD is the primary analysis: one endpoint vector is calculated per acquisition, and all exploratory comparisons reported below use these acquisition-level scalars. The amplitude endpoints $R_0$, $R_m$, and $A_m$ are expressed in $\mathrm{mm\,s^{-1}}$; $\rho_m$, MPR, $R_{\mathrm{eff}}$, and PR are dimensionless.

For any waveform $x(t)$, define
\begin{equation}
\RMS_t(x)
=
\sqrt{\left\langle x(t)^2\right\rangle_t}.
\label{eq:rms_def}
\end{equation}

\subsection{Total, modal, and residual amplitudes}

\begin{figure*}[t]
\centering
\includegraphics[width=1.0\linewidth]{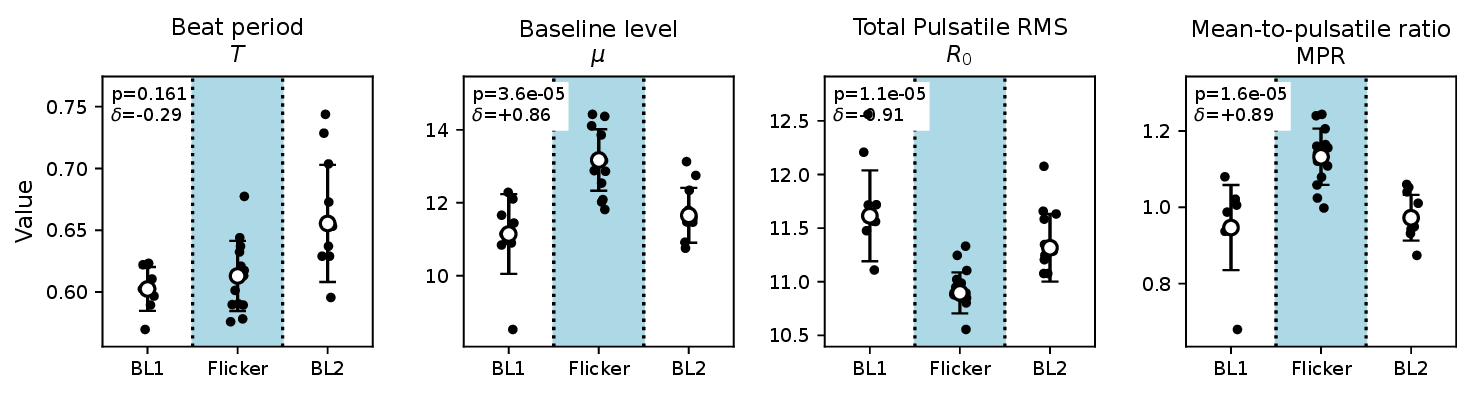}
\caption{Non-SVD arterial context endpoints for the analyzed eye across Baseline~1 (B1), Flicker, and Baseline~2 (B2).
This panel provides the scale and timing context for the low-rank analysis.
Left to right: beat period $T$, baseline velocity level $\mu$, total pulsatile RMS $R_0$ of the locally centered waveform field, and mean-to-pulsatile ratio MPR.
Black dots are individual acquisitions; hollow markers are epoch medians; whiskers are median $\pm$ one sample standard deviation across acquisitions in that epoch.
Acquisition distributions and Cliff's $\delta$ are primary; the displayed raw, unadjusted $p$ values are secondary exploratory summaries of the pooled baseline-versus-flicker comparison
(Flicker vs.\ B1$\cup$B2; $\delta>0$ means Flicker tends larger).
In particular, $T$ and $\mu$ are contextual variables and are not included in the ten-endpoint Holm family.}
\label{fig:nonsvd_endpoint_grid}
\end{figure*}

The residual before removal of any SVD mode is $r_0\equiv w$. Its robust acquisition-level amplitude is
\begin{equation}
\begin{aligned}
R_0
&=\med_{b,k,r}
\RMS_t\!\left[w(t,b,k,r)\right].
\end{aligned}
\label{eq:R0_def}
\end{equation}
Thus, $R_0$ is the robust RMS scale of the locally centered waveform field.

The residual after removal of the first $m$ modes is
\begin{equation}
r_m(t,b,k,r)
=
w(t,b,k,r)
-\sum_{\ell=1}^{m}a_\ell(b,k,r)u_\ell(t),
\label{eq:rm_def}
\end{equation}
with robust residual amplitude
\begin{equation}
R_m
\equiv
\med_{b,k,r}
\RMS_t\!\left[r_m(t,b,k,r)\right].
\label{eq:Rm_def}
\end{equation}
The numerically safeguarded robust residual-amplitude ratio is
\begin{equation}
\rho_m
\equiv
\begin{cases}
\dfrac{R_m}{R_0+\epsilon}, & R_0>\epsilon,\\[4pt]
\text{undefined}, & R_0\leq\epsilon,
\end{cases}
\, m\geq1,
\,
\epsilon=10^{-12}~\mathrm{mm\,s^{-1}},
\label{eq:rhom_def}
\end{equation}
with $\rho_0=1$ whenever $R_0>\epsilon$.
Thus, $\rho_m$ is the ratio between the median residual RMS amplitude after removal of the first $m$ modes and the median RMS amplitude of the original centered waveforms. Accordingly, $\rho_1$ measures the typical residual amplitude after removal of the leading acquisition-specific arterial pulse mode, relative to $R_0$, while $\rho_2$ measures the corresponding relative residual amplitude after removal of the first two modes. These quantities are ratios of robust amplitude summaries, not residual-energy or unexplained-variance fractions and not medians of paired local ratios. Robustness here refers to the median aggregation; the ordinary SVD itself still requires quality control against corrupted waveform columns.

The robust amplitude of mode $m$ is
\begin{equation}
A_m
\equiv
\med_{b,k,r}
\RMS_t\!\left[a_m(b,k,r)u_m(t)\right].
\label{eq:Am_def}
\end{equation}
Because the temporal modes are unit-norm, $A_m$, $R_m$, and $R_0$ share the same signal units. $A_1$ measures expression of the leading acquisition-specific arterial pulse mode, whereas $A_2$ measures the second modal component.

\subsection{Mean-to-pulsatile ratio}

The mean-to-pulsatile ratio is defined as the median of paired local ratios,
\begin{equation}
\mathrm{MPR}
\equiv
\med_{b,k,r}
\left[
\frac{\left|\mu(b,k,r)\right|}
{\RMS_t\!\left[w(t,b,k,r)\right]+\epsilon}
\right].
\label{eq:mpr_def}
\end{equation}
MPR measures the local mean velocity level relative to its pulsatile RMS amplitude. The paired construction preserves the correspondence between the mean and centered components of each waveform. The implementation uses the validity criteria above and the same $\epsilon$ for numerical stabilization; it applies no additional minimum-denominator exclusion.

\subsection{Singular-spectrum endpoints}
\label{sec:spectrum_endpoints}

Let $\sigma_m\equiv\lambda_m^2$ denote the energy of SVD mode $m$, and let $q=\min(N_t,N_j)$ be the number of singular values returned by the economy SVD. The singular-spectrum endpoints are calculated from the complete spectrum, so that $q=N_t$ whenever the number of valid beat--location waveforms satisfies $N_j\geq N_t$. Define
\begin{equation}
p_m
\equiv
\frac{\sigma_m}{\displaystyle\sum_{\ell=1}^{q}\sigma_\ell},
\qquad
m=1,\ldots,q.
\label{eq:normalized_svd_energy}
\end{equation}
The effective rank is the exponential Shannon entropy of the normalized singular-value energy distribution and estimates the effective number of appreciably occupied modes.\cite{Roy2007EffectiveRank} It is defined as
\begin{equation}
R_{\mathrm{eff}}
\equiv
\exp\!\left(-\sum_{m=1}^{q}p_m\log p_m\right),
\label{eq:effective_rank}
\end{equation}
and the participation ratio is
\begin{equation}
\mathrm{PR}
\equiv
\frac{1}{\displaystyle\sum_{m=1}^{q}p_m^2}
=
\frac{\left(\displaystyle\sum_{m=1}^{q}\sigma_m\right)^2}
{\displaystyle\sum_{m=1}^{q}\sigma_m^2}.
\label{eq:participation_ratio}
\end{equation}
Zero-energy terms make no contribution to either endpoint. Both quantities approach one when the centered waveform field is nearly rank one and increase, up to $q$, as variance is distributed across more modes. $R_{\mathrm{eff}}$ is relatively sensitive to a broad tail of small contributions, whereas PR emphasizes modes carrying appreciable variance.

\section{Beat-level endpoints}
\label{sec:beat_level_endpoints}

All endpoints can also be computed from a single beat. The waveform matrix then contains only the centered arterial waveforms from the valid vessel locations of that beat, and an independent SVD is performed. Its temporal modes, scores, and singular values are specific to that beat rather than shared across the acquisition. For beat $b$, $R_{\mathrm{eff}}$ and PR are calculated from all $q_b=\min(N_t,N_{j,b})$ available singular values, where $N_{j,b}$ is the number of valid vessel locations in that beat. For the representative beat, $q_b=\min(128,79)=79$, whereas the primary joint acquisition-level SVD has $q=128$. This beat-level decomposition is a separate descriptive analysis and is not used for the primary acquisition-level comparisons.

\begin{figure}[t]
\centering
\includegraphics[width=0.90\linewidth]{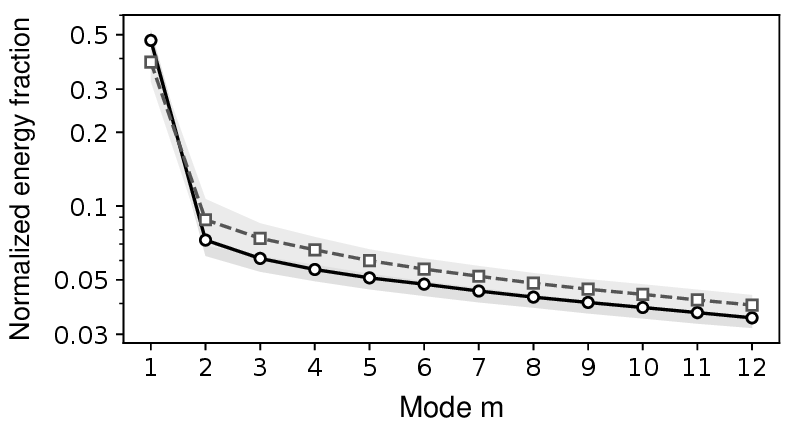}
\caption{Descriptive beat-level analysis of the arterial SVD spectrum in the analyzed eye.
The energy fractions of the first 12 modes, normalized within the displayed set,
$\widetilde p_m^{(12)}=\lambda_m^2/\sum_{\ell=1}^{12}\lambda_\ell^2$,
are shown versus mode index $m$.
The solid black curve with circular markers shows pooled baseline acquisitions;
the dashed gray curve with square markers shows flicker acquisitions.
For each acquisition, the normalized energy fraction at each mode is first averaged across its valid beats; the condition curve is then the mean of these acquisition-level beat means, giving equal weight to each acquisition.
Shaded bands are the condition mean $\pm$ one standard deviation across the individual beat spectra pooled within that condition; the beat is therefore the variability unit underlying the band, although no beat-level inference is performed.
The first 12 fractions are normalized within these 12 displayed modes; their changes are therefore compositional, so an increase in higher-mode shares is coupled to a decrease in the leading-mode share. The 12-mode display limit does not truncate the complete singular spectrum used to calculate beat-level $R_{\mathrm{eff}}$ and PR. This figure is descriptive and is not part of the primary acquisition-level statistical analysis.}
\label{fig:svd_spectrum}
\end{figure}

Figure~\ref{fig:svd_spectrum} displays the normalized energy fractions of the first 12 beat-specific modes for visualization. This plotting range is distinct from the complete $q_b$-mode spectrum used to calculate the singular-spectrum endpoints. Relative to baseline, the flicker spectra show less concentration in the first mode and greater relative contributions from higher displayed modes. Because raw singular values depend on waveform scale and the number of valid locations, the comparison uses normalized modal energies; because those energies are normalized within the displayed set, their changes are compositional.

The definitions in Sec.~\ref{sec:endpoints} are otherwise unchanged, except that $\med_{k,r}$ replaces $\med_{b,k,r}$. There is no aggregation over beats. The amplitude, residual, ratio, MPR, and singular-spectrum endpoints therefore form one arterial endpoint vector for each beat. Any subsequent aggregation across beats is a separate analysis step, and beats from the same acquisition should not be treated as independent biological replicates.

\section{Statistical analysis}
\label{sec:stats}

All comparisons are exploratory within-session comparisons using acquisitions from the analyzed eye as the nominal statistical units. Beats, vessel branches, and radial samples contribute to the robust acquisition-level summaries but are not treated as independent observations.

The observed acquisition distributions, epoch medians, and Cliff's $\delta$ are the primary summaries. As a secondary exploratory calculation, Flicker is compared with the pooled Baseline~1 and Baseline~2 acquisitions using a two-sided Mann--Whitney $U$ test, with Holm correction across the ten-endpoint arterial family. Interpretation emphasizes distributions and effect sizes rather than thresholded significance.

The Mann--Whitney calculation formally treats the 33 temporally ordered acquisitions as independent, although they are repeated measurements from one eye. The resulting raw and Holm-adjusted $p$ values therefore quantify within-session separation under that working assumption; they are not independent biological evidence and do not support subject-level inference.

\section{Results}
\label{sec:results}

\subsection{Arterial endpoint panels}

\begin{figure*}[t]
\centering
\includegraphics[width=1.0\linewidth]{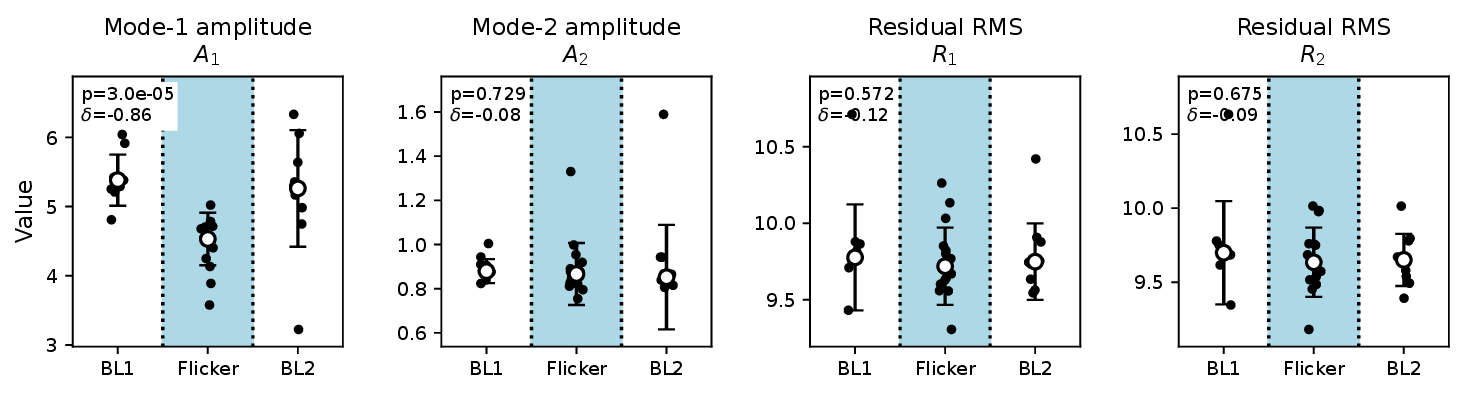}
\caption{Arterial low-rank amplitude and residual endpoints for the analyzed eye across Baseline~1 (B1), Flicker, and Baseline~2 (B2), from joint SVD acquisition scalars.
This panel separates changes in leading-mode amplitude from changes in absolute residual waveform amplitude during flicker stimulation.
Left to right: mode-1 and mode-2 amplitudes $A_1$ and $A_2$, and residual RMS after removal of the first one and two modes, $R_1$ and $R_2$.
Black dots are individual acquisitions; hollow markers are epoch medians; whiskers are median $\pm$ one sample standard deviation across acquisitions.
Cliff's $\delta$ and the secondary exploratory raw $p$ values are for Flicker versus pooled baseline (B1$\cup$B2; $\delta>0$ means Flicker tends larger).}
\label{fig:primary_endpoints}
\end{figure*}

The primary results are derived from the joint acquisition-level SVD. Each acquisition yields one arterial endpoint vector (Fig.~\ref{fig:primary_endpoints}). Relative to the pooled baseline acquisitions, the median total centered-waveform amplitude $R_0$ decreased from $11.6$ to $10.9~\mathrm{mm\,s^{-1}}$ during flicker (Cliff's $\delta=-0.910$), and the leading-mode amplitude $A_1$ decreased from $5.30$ to $4.53~\mathrm{mm\,s^{-1}}$ ($\delta=-0.865$). In parallel, the median residual-amplitude ratios increased from $0.850$ to $0.892$ for $\rho_1$ ($\delta=+0.925$) and from $0.841$ to $0.882$ for $\rho_2$ ($\delta=+0.962$). The return toward the Baseline~1 range during Baseline~2 was descriptively consistent with reversibility. The corresponding exploratory acquisition-level comparisons are reported in Table~\ref{tab:artery_table1_pooled}.

No comparable separation was observed for the second-mode amplitude or the absolute residual amplitudes. The baseline and flicker medians were $0.876$ and $0.867~\mathrm{mm\,s^{-1}}$ for $A_2$ ($\delta=-0.075$), $9.76$ and $9.72~\mathrm{mm\,s^{-1}}$ for $R_1$ ($\delta=-0.120$), and $9.69$ and $9.63~\mathrm{mm\,s^{-1}}$ for $R_2$ ($\delta=-0.090$). Thus, the larger $\rho_m$ values accompanied a reduction in $R_0$, not an increase in absolute residual amplitude.

\subsection{Reduced modal concentration}

In the arterial panel, the increase in $\rho_2$ indicates a larger typical residual RMS amplitude, relative to $R_0$, after removal of the first two modes. The median effective rank increased from $45.7$ at baseline to $54.4$ during flicker ($\delta=+0.827$), while PR increased from $13.0$ to $20.8$ ($\delta=+0.902$). These changes provide complementary evidence of reduced concentration of normalized energy in the leading mode.

\subsection{Interpretation and controls}

Within the arterial data from this eye, the combined behavior of $A_1$, $\rho_1$, $\rho_2$, and the spectrum endpoints is more naturally interpreted as reduced concentration in the leading acquisition-specific arterial pulse mode than as the appearance of a new waveform. No evidence of an $A_2$ difference was detected in the pooled baseline-versus-flicker comparison (Cliff's $\delta=-0.075$; raw $p=0.729$; $p_{\mathrm{H}}=1.000$), and Mode~2 should not be interpreted as a reproducible response component.

The median MPR increased from $0.964$ at baseline to $1.13$ during flicker ($\delta=+0.895$). The contextual panels show an upward shift of $\mu$ and a downward shift of beat period $T$ during flicker. The MPR increase was therefore directionally consistent with both a higher local mean-velocity level and the lower pulsatile scale $R_0$; because MPR is a median of paired local ratios, however, it cannot be reconstructed exactly from the separately aggregated $\mu$ and $R_0$ values. Beat period remains a potential physiological confound rather than a member of the endpoint family. The MPR denominator is stabilized by the specified $\epsilon$, without a separate amplitude-based exclusion.

\begin{figure*}[t]
\centering
\includegraphics[width=1.0\linewidth]{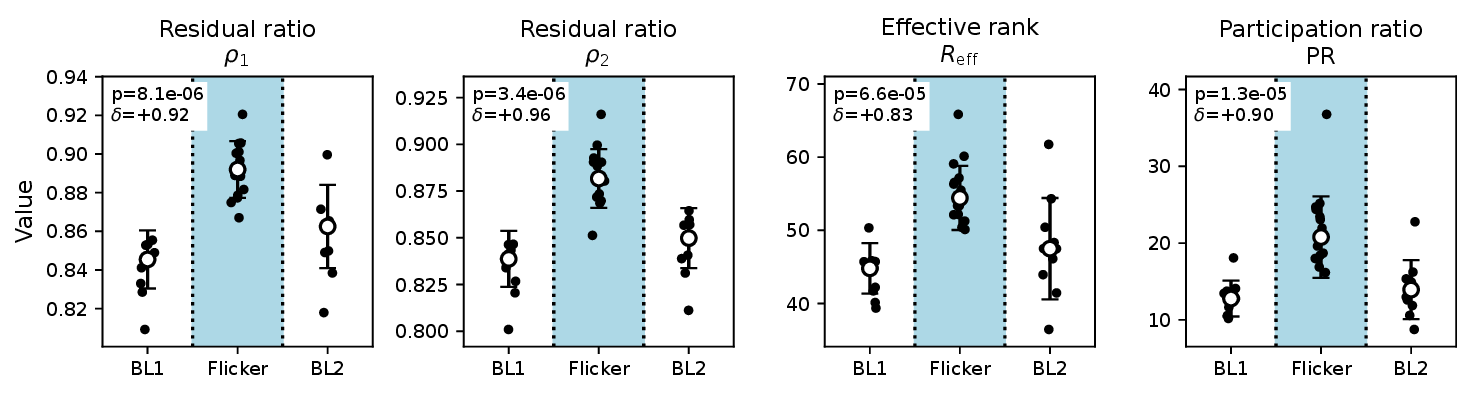}
\caption{Arterial residual-ratio and singular-spectrum endpoints for the analyzed eye across Baseline~1 (B1), Flicker, and Baseline~2 (B2), from joint SVD acquisition scalars.
This panel asks whether flicker changes appear as increased relative residual structure and broader spectral dimensionality, not only as changes in absolute waveform amplitude.
Left to right: residual-amplitude ratios $\rho_1=R_1/R_0$ and $\rho_2=R_2/R_0$, effective rank $R_{\mathrm{eff}}$, and participation ratio PR.
The residual ratios quantify residual RMS after removal of the first one or two modes relative to the total centered-waveform RMS $R_0$;
$R_{\mathrm{eff}}$ and PR summarize how broadly variance is distributed across the singular spectrum.
Black dots are individual acquisitions; hollow markers are epoch medians; whiskers are median $\pm$ one sample standard deviation across acquisitions.
Cliff's $\delta$ and the secondary exploratory raw $p$ values are for Flicker versus pooled baseline (B1$\cup$B2; $\delta>0$ means Flicker tends larger).}
\label{fig:dimensionality_endpoints}
\end{figure*}

Table~\ref{tab:artery_table1_pooled} reports the pooled baseline-versus-flicker comparison for the ten arterial endpoints of the analyzed eye. The principal interpretation centers on $R_0$, $A_1$, $\rho_1$, and $\rho_2$; $A_2$ and the remaining spectrum endpoints provide complementary modal context.

\begin{table*}[t]
\centering
\caption{Pooled baseline (B1+B2) vs.\ flicker comparison for artery endpoints
($n_{\mathrm{baseline}}=19$, $n_{\mathrm{flicker}}=14$).
Values are median [IQR width]. Cliff's $\delta$ is the primary effect-size summary (positive $=$ higher in flicker).
The secondary exploratory $p$ column gives the unadjusted two-sided Mann--Whitney value; $p_{\mathrm{H}}$ is Holm-adjusted within the ten endpoints. These are exploratory within-session comparisons of temporally ordered acquisitions from one eye. A dash in the unit column denotes a dimensionless endpoint.}
\label{tab:artery_table1_pooled}
\begin{tabular*}{0.98\textwidth}{@{\extracolsep{\fill}}lcccccc@{}}
\toprule
Endpoint &
Unit &
Baseline median [IQR width] &
Flicker median [IQR width] &
Cliff's $\delta$ &
$p$ &
$p_{\mathrm{H}}$ \\
\midrule
$A_1$ &
$\mathrm{mm\,s^{-1}}$ &
$5.30\,[0.343]$ &
$4.53\,[0.410]$ &
$-0.865$ &
$3.04\times10^{-5}$ &
$1.52\times10^{-4}$ \\
$A_2$ &
$\mathrm{mm\,s^{-1}}$ &
$0.876\,[0.0905]$ &
$0.867\,[0.0902]$ &
$-0.075$ &
$0.729$ &
$1.000$ \\
$R_0$ &
$\mathrm{mm\,s^{-1}}$ &
$11.6\,[0.420]$ &
$10.9\,[0.156]$ &
$-0.910$ &
$1.14\times10^{-5}$ &
$9.10\times10^{-5}$ \\
$R_1$ &
$\mathrm{mm\,s^{-1}}$ &
$9.76\,[0.150]$ &
$9.72\,[0.235]$ &
$-0.120$ &
$0.572$ &
$1.000$ \\
$R_2$ &
$\mathrm{mm\,s^{-1}}$ &
$9.69\,[0.167]$ &
$9.63\,[0.239]$ &
$-0.090$ &
$0.675$ &
$1.000$ \\
$\rho_1$ &
-- &
$0.850\,[0.0227]$ &
$0.892\,[0.0216]$ &
$0.925$ &
$8.11\times10^{-6}$ &
$7.30\times10^{-5}$ \\
$\rho_2$ &
-- &
$0.841\,[0.0173]$ &
$0.882\,[0.0183]$ &
$0.962$ &
$3.41\times10^{-6}$ &
$3.41\times10^{-5}$ \\
MPR &
-- &
$0.964\,[0.0737]$ &
$1.13\,[0.0769]$ &
$0.895$ &
$1.59\times10^{-5}$ &
$9.51\times10^{-5}$ \\
$R_{\mathrm{eff}}$ &
-- &
$45.7\,[5.95]$ &
$54.4\,[4.86]$ &
$0.827$ &
$6.65\times10^{-5}$ &
$2.66\times10^{-4}$ \\
PR &
-- &
$13.0\,[3.26]$ &
$20.8\,[6.04]$ &
$0.902$ &
$1.34\times10^{-5}$ &
$9.40\times10^{-5}$ \\
\bottomrule
\end{tabular*}
\end{table*}

\section{Discussion}
\label{sec:discussion}

This study shows that flicker stimulation is associated not only with lower retinal arterial pulse-waveform amplitude but also with reduced modal concentration. Across acquisitions from the analyzed eye, the flicker epoch had lower total centered-waveform amplitude $R_0$ and leading-mode amplitude $A_1$, while no evidence of differences in $A_2$ or the absolute residual amplitudes $R_1$ and $R_2$ was detected. At the same time, the robust residual-amplitude ratios $\rho_1$ and $\rho_2$, $R_{\mathrm{eff}}$, and PR were higher. Because no increase in $R_1$ or $R_2$ was detected as $R_0$ decreased, the larger $\rho_m$ values should not be interpreted as greater absolute residual pulsatility. Rather, the combined endpoint pattern indicates reduced concentration of the centered waveform field in its leading acquisition-specific arterial pulse mode and a broader relative distribution across the remaining normalized modal spectrum. The absence of a reproducible $A_2$ increase further argues against attributing the response to the emergence of a single new secondary waveform.

This distinction extends retinal neurovascular-coupling assessment beyond conventional changes in diameter, mean velocity, or total flow. The cardiac waveform reflects the interaction of pulsatile forcing with vascular tone, compliance, resistance, downstream runoff, and spatial heterogeneity. Flicker may therefore change pulsatile scale and modal concentration even when an analysis based only on a mean or peak response is incomplete. The present decomposition does not identify which vascular mechanism produced the observed endpoint pattern, but it quantifies that pattern without assigning causal physiological specificity. Similarly, the increase in MPR indicates a shift in the balance between the local temporal mean and pulsatile RMS amplitude; because MPR is a ratio, it should not by itself be interpreted as evidence that mean velocity increased.

The endpoint families deliberately separate scale from modal concentration. $R_0$, $A_1$, $A_2$, $R_1$, and $R_2$ retain absolute velocity scale and transform linearly under a common positive multiplicative gain $v\mapsto gv$. By contrast, $\rho_1$, $\rho_2$, MPR, $R_{\mathrm{eff}}$, and PR are invariant to that common gain. These normalized endpoints complement previously proposed gain-invariant waveform-shape metrics and are intended to facilitate transportability when absolute calibration differs across sessions, instruments, or sites.\cite{Buisson2026Transportable} Formal gain invariance nevertheless does not confer invariance to additive offsets, clipping, noise, waveform-dependent gains, or changes in the selected beat--location population. Absolute-amplitude comparisons therefore require calibrated or harmonized velocity measurements, while all endpoints require consistent preprocessing and quality control.

The construction has several practical strengths. It uses all beat--location waveforms meeting explicit code-defined validity rules rather than selecting one vessel, one beat, or a manually identified systolic feature. Median endpoint aggregation reduces the influence of isolated extreme scores or residual amplitudes, and the separation of absolute, normalized-residual, and spectral endpoints makes the origin of a change more auditable. The SVD itself, however, is an ordinary least-squares decomposition and is not intrinsically robust to corrupted waveform columns. In particular, the 95\%-finite rule does not detect finite motion artifacts, and the current implementation has no separate residual-motion threshold; upstream motion suppression and acquisition-level quality review remain necessary.

Acquisition-level and beat-level decompositions provide complementary views. The joint acquisition-level SVD estimates the pulse patterns shared across all valid beats and locations and yields one endpoint vector per acquisition, which is the statistical unit used here. A beat-level SVD instead adapts its modes and singular values to the locations available within one beat and can support time-resolved analyses of response dynamics. Because these bases are estimated independently, their modes are not automatically aligned across beats or acquisitions; direct mode-shape comparisons require an explicit alignment or subspace-stability procedure. Beats from the same acquisition also remain correlated and must not be treated as independent biological replicates.

The resulting scalars are automatically measurable, traceable to their component waveforms, and compatible with prespecified validity rules and repeated-measures models. These properties are attractive for longitudinal and multicenter studies of disorders in which flicker responses are impaired or dysregulated, including diabetes, glaucoma, cerebral small-vessel disease, and Alzheimer-spectrum disorders.\cite{Garhofer2004Diabetes,Lott2013Prediabetes,Gugleta2013Glaucoma,Bettermann2012WhiteMatter,Querques2019Alzheimer,Kotliar2017Alzheimer} In that setting, modal endpoints could complement established vascular-reactivity measurements by distinguishing loss of pulsatile scale from reduced modal concentration. The present data show within-session separation in one eye, not diagnostic specificity or clinical utility.

Several limitations constrain interpretation. All acquisitions came from one eye of one healthy participant, and repeated acquisitions within a session cannot substitute for independent subjects. The exploratory Mann--Whitney tests formally treat the 33 temporally ordered acquisitions as independent; their raw and Holm-adjusted $p$ values quantify within-session separation under that working assumption, not independent biological evidence. The fixed Baseline~1/Flicker/Baseline~2 order also means that stimulation effects cannot be completely separated from order or time effects. The return toward the Baseline~1 range during Baseline~2 was descriptively consistent with reversibility, but it does not establish causality. Finally, the SVD basis is data-dependent: modes can rotate or exchange order when singular values are close. Because the spectral endpoints use the complete economy-SVD spectrum, they can also be influenced by the resampling length $N_t$, by the number of valid waveform columns when $N_j<N_t$, and by a tail of small noise-dominated singular values.

Validation should therefore lock the endpoint panel, within-beat resampling length $N_t$, complete-spectrum convention, waveform-validity criteria, and preprocessing pipeline before cohort analysis. Multi-subject and multi-day studies should quantify test--retest repeatability, between-eye and between-subject variability, sensitivity to beat period, vessel selection and signal quality, and robustness across instruments and sites.\cite{Kalitzeos2026Repeatability} Such studies should use hierarchical or repeated-measures inference at the participant level and determine which absolute and gain-invariant endpoints preserve the observed provocation sensitivity while meeting the reproducibility requirements of clinical research.

\section{Conclusion}

Doppler holography provides a beat-resolved arterial waveform dimension of retinal neurovascular coupling that is not captured by conventional diameter, mean-velocity, flow, or perfusion-density endpoints. Acquisition-specific low-rank decomposition converts this high-dimensional information into a compact and auditable panel that separates absolute pulsatile scale, leading-mode expression, relative residual structure, mean-to-pulsatile balance, and singular-spectrum dimensionality.

In the analyzed eye, the flicker epoch had lower $R_0$ and $A_1$, no evidence of differences in $R_1$, $R_2$, or $A_2$ was detected, and $\rho_1$, $\rho_2$, $R_{\mathrm{eff}}$, and PR were higher. The observed separation is therefore most consistently described as attenuation of the leading acquisition-specific arterial pulse mode and reduced modal concentration, rather than amplification of absolute residual pulsatility or emergence of a reproducible second waveform. This interpretation illustrates the complementary value of reporting absolute amplitudes together with normalized residual and spectral endpoints.

The endpoints can be generated automatically from each acquisition or beat and include gain-invariant quantities designed for transportability, making the framework suitable for standardized longitudinal studies and patient cohorts. The present one-eye experiment demonstrates computational feasibility and within-session separation but does not establish biological reproducibility, causal neurovascular specificity, clinical validity, or a population biomarker. Multi-subject, multi-day, and multicenter test--retest studies are now required to determine which endpoints preserve this separation while achieving the reproducibility, calibration stability, and disease specificity needed for clinical assessment of retinal neurovascular function.

\section*{Data and software availability}

The dataset supporting this study is openly available on Zenodo at
\href{https://doi.org/10.5281/zenodo.21938682}{doi:10.5281/zenodo.21938682}.
The corresponding processing and analysis software is available in the tagged releases
\href{https://github.com/DigitalHolography/EyeFlow/releases/tag/v1.15.0}{EyeFlow v1.15.0}
(commit \href{https://github.com/DigitalHolography/EyeFlow/commit/064f54705fc9428a43c4a9a80423a5bd146407b5}{\texttt{064f547}})
and
\href{https://github.com/DigitalHolography/AngioEye/releases/tag/v1.23.0}{AngioEye v1.23.0}
(commit \href{https://github.com/DigitalHolography/AngioEye/commit/1d9607fab222add2d334128d3ceb15bed0c94fbf}{\texttt{1d9607f}}).

\bibliographystyle{apsrev4-2}
\bibliography{SVD_endpoints}

@article{Garhofer2004FlickerFlow,
  author  = {Garh{\"o}fer, Gerhard and Zawinka, Christian and Resch, Herbert and Huemer, Karl Heinz and Dorner, Gerhard T. and Schmetterer, Leopold},
  title   = {Diffuse luminance flicker increases blood flow in major retinal arteries and veins},
  journal = {Vision Research},
  year    = {2004},
  volume  = {44},
  number  = {8},
  pages   = {833--838},
  doi     = {10.1016/j.visres.2003.11.013}
}

@article{Wang2011DopplerOCT,
  author  = {Wang, Yimin and Fawzi, Amani A. and Tan, Ou and Zhang, Xinbo and Huang, David},
  title   = {Flicker-induced changes in retinal blood flow assessed by {Doppler} optical coherence tomography},
  journal = {Biomedical Optics Express},
  year    = {2011},
  volume  = {2},
  number  = {7},
  pages   = {1852--1860},
  doi     = {10.1364/BOE.2.001852}
}

@article{Aschinger2017TotalFlow,
  author  = {Aschinger, Gerold C. and Schmetterer, Leopold and Fondi, Klemens and Aranha Dos Santos, Valentin and Seidel, Gerald and Garh{\"o}fer, Gerhard and Werkmeister, Ren{\'e} M.},
  title   = {Effect of diffuse luminance flicker light stimulation on total retinal blood flow assessed with dual-beam bidirectional {Doppler OCT}},
  journal = {Investigative Ophthalmology \& Visual Science},
  year    = {2017},
  volume  = {58},
  number  = {2},
  pages   = {1167--1178},
  doi     = {10.1167/iovs.16-20598}
}

@article{Warner2020ParafovealFlow,
  author  = {Warner, Raymond L. and de Castro, Alberto and Sawides, Lucie and Gast, Tom and Sapoznik, Kaitlyn and Luo, Ting and Burns, Stephen A.},
  title   = {Full-field flicker evoked changes in parafoveal retinal blood flow},
  journal = {Scientific Reports},
  year    = {2020},
  volume  = {10},
  pages   = {16051},
  doi     = {10.1038/s41598-020-73032-0}
}

@article{Kallab2021PlexusSpecific,
  author  = {Kallab, Martin and Hommer, Nikolaus and Tan, Bingyao and Pfister, Martin and Schlatter, Andreas and Werkmeister, Ren{\'e} M. and Chua, Jacqueline and Schmidl, Doreen and Schmetterer, Leopold and Garh{\"o}fer, Gerhard},
  title   = {Plexus-specific effect of flicker-light stimulation on the retinal microvasculature assessed with optical coherence tomography angiography},
  journal = {American Journal of Physiology-Heart and Circulatory Physiology},
  year    = {2021},
  volume  = {320},
  number  = {1},
  pages   = {H23--H28},
  doi     = {10.1152/ajpheart.00495.2020}
}

@article{Huang2025CircularOCTA,
  author  = {Huang, Naixing and Hormel, Tristan T. and Chen, Siyu and Huang, David and Hwang, Thomas S. and Bailey, Steven T. and Jia, Yali},
  title   = {Retinal neurovascular coupling evaluation with circular-scan {OCTA}},
  journal = {Optics Express},
  year    = {2025},
  volume  = {33},
  number  = {15},
  pages   = {31295--31306},
  doi     = {10.1364/OE.567888}
}

@article{Kalitzeos2026Repeatability,
  author  = {Kalitzeos, Angelos and Summers, Robert J. and Heitmar, Rebekka},
  title   = {Repeatability of retinal vessel flicker responses in healthy individuals},
  journal = {Acta Ophthalmologica},
  year    = {2026},
  volume  = {104},
  number  = {2},
  pages   = {e165--e172},
  doi     = {10.1111/aos.17578}
}

@article{Puyo2018InVivoLDH,
  author  = {Puyo, L{\'e}o and Paques, Michel and Fink, Mathias and Sahel, Jos{\'e}-Alain and Atlan, Michael},
  title   = {In vivo laser {Doppler} holography of the human retina},
  journal = {Biomedical Optics Express},
  year    = {2018},
  volume  = {9},
  number  = {9},
  pages   = {4113--4129},
  doi     = {10.1364/BOE.9.004113}
}

@article{Puyo2019Waveform,
  author  = {Puyo, L{\'e}o and Paques, Michel and Fink, Mathias and Sahel, Jos{\'e}-Alain and Atlan, Michael},
  title   = {Waveform analysis of human retinal and choroidal blood flow with laser {Doppler} holography},
  journal = {Biomedical Optics Express},
  year    = {2019},
  volume  = {10},
  number  = {10},
  pages   = {4942--4963},
  doi     = {10.1364/BOE.10.004942}
}

@article{Gugleta2013Glaucoma,
  author  = {Gugleta, K. and Waldmann, N. and Polunina, A. and Kochkorov, A. and Katamay, R. and Flammer, J. and Org{\"u}l, S.},
  title   = {Retinal neurovascular coupling in patients with glaucoma and ocular hypertension and its association with the level of glaucomatous damage},
  journal = {Graefe's Archive for Clinical and Experimental Ophthalmology},
  year    = {2013},
  volume  = {251},
  number  = {6},
  pages   = {1577--1585},
  doi     = {10.1007/s00417-013-2276-9}
}

@article{Newman2013,
  author  = {Newman, Eric A.},
  title   = {Functional hyperemia and mechanisms of neurovascular coupling in the retinal vasculature},
  journal = {Journal of Cerebral Blood Flow \& Metabolism},
  year    = {2013},
  volume  = {33},
  number  = {11},
  pages   = {1685--1695},
  doi     = {10.1038/jcbfm.2013.145}
}

@article{Aung2024,
  author  = {Aung, Moe H. and Aleman, Tomas S. and Garcia, Arielle S. and McGeehan, Brendan and Ying, Gui-Shuang and Avery, Robert A.},
  title   = {Stimulus type and duration affect magnitude and evolution of flicker-induced hyperemia measured by laser speckle flowgraphy at the optic disc and peripapillary vessels},
  journal = {Scientific Reports},
  year    = {2024},
  volume  = {14},
  pages   = {6659},
  doi     = {10.1038/s41598-024-57263-z}
}

@article{Kwan2020,
  author  = {Kwan, C. C. and Lee, H. E. and Schwartz, G. and Fawzi, A. A.},
  title   = {Acute hyperglycemia reverses neurovascular coupling during dark to light adaptation in healthy subjects on optical coherence tomography angiography},
  journal = {Investigative Ophthalmology \& Visual Science},
  year    = {2020},
  volume  = {61},
  number  = {4},
  pages   = {38},
  doi     = {10.1167/iovs.61.4.38}
}

@article{Liu2023fOCTA,
  author  = {Liu, Kaiyuan and Zhu, Tiepei and Gao, Mengqin and Yin, Xiaoting and Zheng, Rong and Yan, Yan and Gao, Lei and Ding, Zhihua and Ye, Juan and Li, Peng},
  title   = {Functional {OCT} angiography reveals early retinal neurovascular dysfunction in diabetes with capillary resolution},
  journal = {Biomedical Optics Express},
  year    = {2023},
  volume  = {14},
  number  = {4},
  pages   = {1670--1684},
  doi     = {10.1364/BOE.485940}
}

@article{Peterfi2024,
  author  = {Peterfi, Anna and others},
  title   = {Dynamic retinal vessel analysis: flickering a light into the brain},
  journal = {Frontiers in Aging Neuroscience},
  year    = {2024},
  volume  = {16},
  pages   = {1517368},
  doi     = {10.3389/fnagi.2024.1517368}
}

@article{Senee2025SciAdv,
  author  = {Sen{\'e}e, Pierre and Krafft, L{\'e}a and Loukili, In{\`e}s and Castro Farias, Daniela and Thouvenin, Olivier and Atlan, Michael and Paques, Michel and Meimon, Serge and Mec{\^e}, Pedro},
  title   = {Revealing neurovascular coupling at a high spatial and temporal resolution in the living human retina},
  journal = {Science Advances},
  year    = {2025},
  volume  = {11},
  pages   = {eadx2941},
  doi     = {10.1126/sciadv.adx2941}
}

@misc{Bratasz2022Diffuse,
  author        = {Bratasz, Zofia and Martinache, Olivier and Blazy, Yohan and Denis, Ang{\`e}le and Auffret, Coline and Huignard, Jean-Pierre and Rossi, Ethan and Chhablani, Jay and Sahel, Jos{\'e}-Alain and Bonnin, Sophie and Hage, Rabih and Koskas, Patricia and Gatinel, Damien and Vignal, Catherine and Yavchitz, Am{\'e}lie and Vasseur, Vivien and Tadayoni, Ramin and Ducloux, Claire and Ortoli, Manon and Tordjman, Marvin and Tick, Sarah and Mrejen, Sarah and Paques, Michel and Atlan, Michael},
  title         = {Diffuse {Maxwellian} illumination for safe wide-field retinal {Doppler} holography},
  year          = {2022},
  eprint        = {2212.13347},
  archivePrefix = {arXiv},
  primaryClass  = {physics.optics},
  doi           = {10.48550/arXiv.2212.13347},
  note          = {Version 3, revised May 2026}
}

@misc{Doucet2026LowRank,
  author    = {Doucet, Lancelot and Gouyoumdjian, David and Boy Arnould, Maxime and Fischer, Yann and Auray, Zacharie and Atlan, Michael},
  title     = {Low-Rank Modal Decomposition of Beat-Resolved Retinal Laser Doppler Holography at 37,000 Frames per Second Reveals Flicker-Evoked Arterial Neurovascular Signatures},
  year      = {2026},
  publisher = {Zenodo},
  doi       = {10.5281/zenodo.18601591},
  url       = {https://doi.org/10.5281/zenodo.18601591}
}

@misc{Buisson2026Transportable,
  author    = {Buisson, Gr{\'e}goire and Paquet, Chlo{\'e} and Doucet, Lancelot and Gouyoumdjian, David and Boy Arnould, Maxime and Fischer, Yann and Auray, Zacharie and Atlan, Michael},
  title     = {Transportable retinal {Doppler} holography waveform-shape metrics for functional microvascular assessment},
  year      = {2026},
  publisher = {Zenodo},
  doi       = {10.5281/zenodo.18895922},
  url       = {https://doi.org/10.5281/zenodo.18895922}
}

@misc{Auray2025UltrahighSpeed,
  author        = {Auray, Zacharie and Fischer, Yann and Martinache, Olivier and Atlan, Michael},
  title         = {Ultrahigh-speed digital holography for quantitative {Doppler} imaging of the human retina},
  year          = {2025},
  eprint        = {2505.07823},
  archivePrefix = {arXiv},
  primaryClass  = {physics.ins-det},
  doi           = {10.48550/arXiv.2505.07823}
}

@misc{Fischer2024RetinalFlow,
  author        = {Fischer, Yann and Auray, Zacharie and Martinache, Olivier and Dubosc, Marius and Top{\'e}za, No{\'e} and Magnier, Chlo{\'e} and Boy-Arnould, Maxime and Atlan, Michael},
  title         = {Retinal arterial blood flow measured by real-time {Doppler} holography at 33,000 frames per second},
  year          = {2024},
  eprint        = {2409.17180},
  archivePrefix = {arXiv},
  primaryClass  = {eess.IV},
  doi           = {10.48550/arXiv.2409.17180}
}

@article{Garhofer2004Diabetes,
  author  = {Garh{\"o}fer, Gerhard and Zawinka, Christian and Resch, Herbert and Kothy, P. and Schmetterer, Leopold and Dorner, Gerhard T.},
  title   = {Reduced Response of Retinal Vessel Diameters to Flicker Stimulation in Patients with Diabetes},
  journal = {British Journal of Ophthalmology},
  year    = {2004},
  volume  = {88},
  number  = {7},
  pages   = {887--891},
  doi     = {10.1136/bjo.2003.033548}
}

@article{Mandecka2007Diabetes,
  author  = {Mandecka, Aleksandra and Dawczynski, Jens and Blum, Marcus and M{\"u}ller, Nicolle and Kloos, Christoph and Wolf, Gunter and Vilser, Walthard and Hoyer, Heike and M{\"u}ller, Ulrich Alfons},
  title   = {Influence of Flickering Light on the Retinal Vessels in Diabetic Patients},
  journal = {Diabetes Care},
  year    = {2007},
  volume  = {30},
  number  = {12},
  pages   = {3048--3052},
  doi     = {10.2337/dc07-0927}
}

@article{Lott2013Prediabetes,
  author  = {Lott, Mary E. J. and Slocomb, Julia E. and Shivkumar, Vikram and Smith, Bruce and Quillen, David and Gabbay, Robert A. and Gardner, Thomas W. and Bettermann, Kerstin},
  title   = {Impaired Retinal Vasodilator Responses in Prediabetes and Type 2 Diabetes},
  journal = {Acta Ophthalmologica},
  year    = {2013},
  volume  = {91},
  number  = {6},
  pages   = {e462--e469},
  doi     = {10.1111/aos.12129}
}

@article{Lim2017DRProgression,
  author  = {Lim, Laurence S. and Ling, Lieng H. and Ong, Peng Guan and Foulds, Wallace and Tai, E. Shyong and Wong, Tien Yin},
  title   = {Dynamic Responses in Retinal Vessel Caliber With Flicker Light Stimulation and Risk of Diabetic Retinopathy and Its Progression},
  journal = {Investigative Ophthalmology \& Visual Science},
  year    = {2017},
  volume  = {58},
  number  = {5},
  pages   = {2449--2455},
  doi     = {10.1167/iovs.16-21008}
}

@article{Pemp2009Endothelial,
  author  = {Pemp, Berthold and Weigert, Guenther and Karl, Katharina and Petzl, Ursula and Wolzt, Michael and Schmetterer, Leopold and Garhofer, Gerhard},
  title   = {Correlation of Flicker-Induced and Flow-Mediated Vasodilatation in Patients With Endothelial Dysfunction and Healthy Volunteers},
  journal = {Diabetes Care},
  year    = {2009},
  volume  = {32},
  number  = {8},
  pages   = {1536--1541},
  doi     = {10.2337/dc08-2130}
}

@article{Bettermann2012WhiteMatter,
  author  = {Bettermann, Kerstin and Slocomb, Julia E. and Shivkumar, Vikram and Lott, Mary E. J.},
  title   = {Retinal Vasoreactivity as a Marker for Chronic Ischemic White Matter Disease?},
  journal = {Journal of the Neurological Sciences},
  year    = {2012},
  volume  = {322},
  number  = {1--2},
  pages   = {206--210},
  doi     = {10.1016/j.jns.2012.05.041}
}

@article{Blair2025cSVD,
  author  = {Blair, Gordon W. and MacCormick, Ian J. and McGrory, Sarah and MacGillivray, Tom and Hamilton, Iona and Shi, Yulu and Chappell, Francesca and Thrippleton, Michael J. and Stringer, Michael S. and Doubal, Fergus and Wardlaw, Joanna M.},
  title   = {Retinal and Cerebral Vascular Reactivity in Cerebral Small Vessel Disease},
  journal = {Journal of Cerebral Blood Flow \& Metabolism},
  year    = {2025},
  volume  = {45},
  number  = {12},
  doi     = {10.1177/0271678X251366079}
}

@article{Querques2019Alzheimer,
  author  = {Querques, Giuseppe and Borrelli, Enrico and Sacconi, Riccardo and De Vitis, Luigi and Leocani, Letizia and Santangelo, Roberto and Magnani, Giuseppe and Comi, Giancarlo and Bandello, Francesco},
  title   = {Functional and Morphological Changes of the Retinal Vessels in Alzheimer's Disease and Mild Cognitive Impairment},
  journal = {Scientific Reports},
  year    = {2019},
  volume  = {9},
  pages   = {63},
  doi     = {10.1038/s41598-018-37271-6}
}

@article{Kotliar2017Alzheimer,
  author  = {Kotliar, Konstantin and Hauser, Christine and Ortner, Marion and Muggenthaler, Claudia and Diehl-Schmid, Janine and Angermann, Susanne and Hapfelmeier, Alexander and Schmaderer, Christoph and Grimmer, Timo},
  title   = {Altered Neurovascular Coupling as Measured by Optical Imaging: A Biomarker for Alzheimer's Disease},
  journal = {Scientific Reports},
  year    = {2017},
  volume  = {7},
  pages   = {12906},
  doi     = {10.1038/s41598-017-13349-5}
}

@inproceedings{Dubosc2026Segmentation,
  author        = {Dubosc, Marius and Fischer, Yann and Auray, Zacharie and Boutry, Nicolas and Carlinet, Edwin and Atlan, Michael and G{\'e}raud, Thierry},
  title         = {Improving Segmentation of Retinal Arteries and Veins Using Cardiac Signal in {Doppler} Holograms},
  booktitle     = {2026 {IEEE} 23rd International Symposium on Biomedical Imaging ({ISBI})},
  year          = {2026},
  pages         = {1--5},
  doi           = {10.1109/ISBI61048.2026.11515426},
  eprint        = {2511.14654},
  archivePrefix = {arXiv},
  primaryClass  = {cs.CV}
}

@inproceedings{Roy2007EffectiveRank,
  author    = {Roy, Olivier and Vetterli, Martin},
  title     = {The Effective Rank: A Measure of Effective Dimensionality},
  booktitle = {15th European Signal Processing Conference ({EUSIPCO})},
  year      = {2007},
  pages     = {606--610},
  address   = {Pozna{\'n}, Poland}
}

\end{document}